\documentclass[a4paper,12pt]{article}
\usepackage{natbib}
\usepackage{hyperref}
\usepackage{epsf}
\usepackage{epsfig}
\usepackage{multirow}
\usepackage[english]{babel}
\usepackage{graphicx}
\usepackage{geometry}
\usepackage{subfigure}
\usepackage{natbib}
\usepackage{amssymb}
\usepackage{amsmath}
\usepackage{color}
\usepackage{enumerate}
\usepackage{authblk}

\title{\textbf{\textbf{MetSizeR}: selecting the optimal sample size for metabolomic studies using an analysis based approach}}

\begin{document}
\pagestyle{plain}

\author[1]{Gift Nyamundanda}
\author[1]{Isobel Claire Gormley\thanks{claire.gormley@ucd.ie}}
\author[2]{Yue Fan}
\author[2]{William M Gallagher}
\author[3]{Lorraine Brennan}
\affil[1]{{\footnotesize School of Mathematical Sciences, University College Dublin, Ireland.}}
\affil[2]{{\footnotesize School of Biomolecular and Biomedical Science, University College Dublin, Ireland.}}
\affil[3]{{\footnotesize School of Agriculture and Food Science, Conway Institute, University
College Dublin, Ireland.}}

\maketitle

\begin{abstract}
Background: Determining sample sizes for metabolomic experiments is important but due to the complexity of these experiments, there are currently no standard methods for sample size estimation in metabolomics. Since pilot studies are rarely done in metabolomics, currently existing sample size estimation approaches which rely on pilot data can not be applied.

Results: In this article, an analysis based approach called MetSizeR is developed to estimate sample size for metabolomic experiments even when experimental pilot data are not available. The key motivation for MetSizeR is that it considers the type of analysis the researcher intends to use for data analysis when estimating sample size. MetSizeR uses information about the data analysis technique and prior expert knowledge of the metabolomic experiment to simulate pilot data from a statistical model. Permutation based techniques are then applied to the simulated pilot data to estimate the required sample size.

Conclusions: The MetSizeR methodology, and a publicly available software package which implements the approach, are illustrated through real metabolomic applications. Sample size estimates, informed by the intended statistical analysis technique, and the associated uncertainty are provided.
\end{abstract}

\section{Background}

In many metabolomic experiments, one of the most important objectives is to discover the set of metabolites that plays a significant role in distinguishing samples from two different groups or populations and thus, in the identification of novel biomarkers \cite{Berk11}. As in any experiment, designing the experiment is critical if reliable statistically significant metabolites are to be obtained. Since metabolomic experiments are expensive, it is crucial to determine the optimal sample size $\hat{n}$ to attain the desired power to identify discriminating metabolites without wasting resources or unnecessarily sampling many subjects. However, metabolomic data are typically high dimensional and correlated meaning sample size estimation using classical statistical methods is not straight forward. \\

Currently, in the metabolomics literature, there is no standard method for the determination of sample size when designing a metabolomic experiment. Several methods currently exist in the literature for sample size selection in the high-dimensional data setting \cite{muller04, tibshirani06, Liu07, lin10}. However, none of these methods are suitable for metabolomic experiments since many either assume variables have equal variance or are independent. More importantly, these methods rely on the existence of some experimental pilot data on which the actual sample size selection is then based, and are not based on the method to be used to analyze the data. In metabolomic studies, experimental pilot data are rarely available, making such sample size selection approaches redundant.

In this article, we propose a method known as MetSizeR for sample size estimation for metabolomic experiments that addresses some of these limitations. MetSizeR is founded on the idea that the method for selecting sample size firmly depends on the type of data analysis the researcher intends to employ. In a situation where experimental pilot data are not available, pseudo-metabolomic data are simulated from a statistical model. The specific statistical model from which the pseudo-metabolomic data are simulated depends on the type of statistical analysis that the metabolomic scientist intends to use. In its current form the MetSizeR approach assumes the user intends to employ one of the following three statistical analysis techniques on completion of their experiment:
\begin{enumerate}
\item Probabilistic Principal Components Analysis (PPCA) \cite{tipping99, nyamundanda10}.
\item Probabilistic Principal Components and Covariates Analysis (PPCCA) \cite{nyamundanda10}.
\item Dynamic Probabilistic Principal Components Analysis (DPPCA) \cite{nyamundanda12a}.
\end{enumerate}
Intuitively the MetSizeR method can be naturally extended to include other analysis approaches, assuming they are based on a statistical model rather than being non-parametric in nature. \\

 MetSizeR draws on two currently existing methods (see \cite{muller04} and \cite{tibshirani06}) for sample size calculation in high-dimensional data settings. While the approach in \cite{tibshirani06} is based on the existence of an experimental pilot data set, the approach detailed in \cite{muller04} simulates pilot data from a statistical model. Further, while independence in the data is assumed in \cite{muller04}, the approach in \cite{tibshirani06} uses permutation methods to account for the correlation in the experimental pilot data. MetSizeR combines these ideas of prior simulation and permutation based techniques to estimate the sample size for metabolomic experiments. The main advantage of the developed approach is its ability to determine sample size without experimental pilot data and without assuming variable independence.\\

A graphic user interface (GUI) software called MetSizeR was developed to implement this approach to estimating sample sizes in R \cite{R09}. Effort was focused on designing the interface of MetSizeR to encourage its wide use in the metabolomics community regardless of previous knowledge of R. The software is available through the \textbf{R} statistical software environment \texttt{www.r-project.org}.

\section{Methods}

Metabolomic data sets are typically acquired using analytical technologies such as nuclear magnetic resonance spectroscopy (NMR) \cite{Reo2002} and mass spectrometry (MS) \cite{Dettmer07}. The spectrum resulting from NMR spectroscopy is usually divided into spectral bins (representing variables) and the signal intensities within the bins are related to the relative abundances of metabolites. MS is typically used for targeted metabolomics in which a specified list of metabolites is measured \cite{Patti12}. The following section describes how the number of samples required for either an NMR or an MS metabolomic experiment can be determined under the MetSizeR approach.

\subsection{Sample size estimation}

Let $\bar{x}_{jg}$ be the estimate of the average signal intensity $\mu_{jg}$ for metabolite $j$ in samples from the treatment group $g$ which has corresponding sample size $n_g$, where $g=1,2$. Often in metabolomics, the goal of discovering a set of metabolites that discriminates between samples from two treatment groups is achieved by testing the hypothesis $\mbox{H}_{oj}:\mu_{j1}-\mu_{j2}=0$, on each metabolite $j$, where $j=1,\ldots,p$. The aim of discovering discriminating metabolites can be framed as a multiple testing problem as there are $p$ hypotheses to be tested and the probability of falsely declaring a metabolite as significant increases with $p$. It is therefore important to estimate sample size while controlling an error rate to improve the power of the test for identifying significant metabolites. MetSizeR focuses on controlling the false discovery rate (FDR, \cite{Benjamini95}). Here, the FDR is the expected number of metabolites incorrectly deemed to be significantly different between the two treatment groups, as a proportion of the total number of metabolites declared to be significant.

\subsubsection*{The test statistic and its distribution}

A test statistic widely used to identify discriminating metabolites is a two sample $t$-statistic. The $t$-statistic $TS$ is evaluated for all metabolites, $j=1, \ldots, p$, under the assumption that the null hypothesis of no difference $\mu_{j1}=\mu_{j2}$ is true:
\begin{eqnarray*}
TS_j&=&\frac{(\bar{x}_{j1}-\bar{x}_{j2})}{S_j + cf},\\
\mbox{where}~~~~~S_j &=& \left\{\left(\frac{1}{n_1}+\frac{1}{n_2}\right)\frac{(n_1-1)(s_{j1})^2+(n_2-1)(s_{j2})^2}{n_1+n_2-2}\right\}^{\frac{1}{2}},
\end{eqnarray*}
where $S_j$ is the estimate of the pooled standard error for metabolite \emph{j}. The corresponding within treatment variability estimate is $s^2_{jg}=(n_g-1)^{-1}\sum_{i=1}^{n_g}(x_{(jg)i}-\bar{x}_{jg})^2$ for $g=1,2$ where $x_{(jg)i}$ denotes the signal intensity for metabolite $j$ in sample $i$ from the treatment group $g$. A correction factor $cf$ is a small positive value added to the standard error of each metabolite to prevent some metabolites with signal intensity near zero from having large test statistic $TS_j$; such a metabolite may have $TS_{j} \approx 0/0$.\\

The typical assumption about the null distribution (i.e. the distribution under the null hypotheses) of the test statistic $TS_j$ is the \emph{t}-distribution with $n_1+n_2-2$ degrees of freedom. However, when the data violate such an assumption, misleading sample size estimates would result. Hence, as in \cite{tibshirani06}, MetSizeR  estimates the null distribution of $TS_j$ using a permutation technique. This is a non-parametric method based on the assumption that under the null hypothesis of no difference, the distribution of the test statistic does not change no matter how the group labels of the pilot data are permuted. The data generated using this approach maintains the between subject variability and the amount of noise in the data. The null distribution of the test statistic $TS$ is
estimated by randomly permuting the group labels of pilot data and calculating the test statistic for each metabolite, $TS_{jt}$, where $t=1, \ldots, T$ permutations.

\subsubsection*{Analysis based data simulation}

Unfortunately, in most cases, experimental pilot data are not readily available in metabolomics since pilot studies are rarely done. Therefore, MetSizeR uses the intended statistical analysis model to simulate pilot data. The simulated pilot data are then used to learn about the null distribution of the relevant test statistic for estimating sample size. This simulation approach is similar to that in \cite{muller04} in which pilot data are simulated from the marginal model:
\begin{eqnarray*}
p(\mathbf{x}) &=& \int p(\mathbf{x}| \mathbf{u}, \theta) dp(\mathbf{u}, \theta),
\label{eqn:margmodel}
\end{eqnarray*}
where $\mathbf{x}$ is the $n \times p$ data matrix, $\mathbf{u}$ denotes the latent variables, and $\theta$ is a vector of unknown model parameters. Simulating from the marginal model is achieved by first generating values of the parameters and the latent variables from the prior distribution $p(\mathbf{u}, \theta)$, and then simulating the data from the assumed model $p(\mathbf{x}| \mathbf{u}, \theta)$ given the simulated values of $\mathbf{u}$ and $\theta$.\\

Currently, MetSizeR assumes the metabolomic practitioner will use one of three different statistical models $p(\mathbf{x}| \mathbf{u}, \theta)$ to analyse the data from their metabolomic experiment -- either the PPCA, PPCCA or DPPCA model. Simulation of the parameters of these models is based on the model assumptions and on prior expert knowledge of metabolomic data properties. As PPCA is equivalent to the widely used Principal Components Analysis (PCA) method, simulating from the PPCA model is discussed here; details of the simulation of pilot data from the closely related PPCCA and DPPCA models are provided in the Additional File.
 Specifically, PPCA is a probabilistic formulation of PCA based on a Gaussian latent variable model \cite{tipping99, nyamundanda10}. PPCA models the high dimensional spectrum $\underline{x}_i^{T}=(x_{i1}, \ldots, x_{ip})$ of subject $i$ ($i=1,\ldots,n$ where $n= n_1 + n_2$) as a linear function of the corresponding low dimensional latent variable $\underline{u}_i^{T}=(u_{i1}, \ldots, u_{iq})$, where $(q \ll p)$. The PPCA model can be expressed as follows
\begin{eqnarray*}
\underline{x}_i & = & \textbf{W} \underline{u}_i  + \underline{\mu} + \underline{\epsilon}_i
\end{eqnarray*}
where $\textbf{W}$ is a $p \times q$ loadings matrix, $\underline{\mu}$ is a mean vector and $\underline{\epsilon}_i$ is multivariate Gaussian noise for observation $i$, i.e. $p(\underline{\epsilon}_i)
\, = \, \mbox{MVN}_{p}(\,\underline{0}\,,\,\sigma^2\textbf{I}\,)$  where \textbf{I} denotes the identity matrix. The latent variable $\underline{u}_i$ is also multivariate Gaussian distributed, $p(\underline{u}_i) \,=  \, \mbox{MVN}_{q}(\,\underline{0}\,,\,\textbf{I}\,)$. The maximum likelihood estimates of the loadings matrix $\mathbf{W}$ and the latent variable $\mathbf{u}$ in the PPCA model are equivalent to the traditional PCA loadings matrix and principal component scores. For a given sample size $n$, pilot data $\mathbf{x} $ can be simulated from the PPCA model as follows:

\begin{enumerate}
 \item Generate parameter values from their prior distributions:
\begin{eqnarray*}
\label{eqn:sig}
p(\underline{u}_{i}) & = & \mbox{MVN}_{q}(\underline{0}, \textbf{I}) \:\: \mbox{for } i =1, \ldots n.\\
p(\underline{w}_{j}) & = & \mbox{MVN}_{q}(\underline{\mu}_{W}, \mathbf{\Sigma}_{W}) \:\: \mbox{for } j = 1, \ldots p.\\
p(\sigma^2) & = & \mbox{IG}(\alpha_1, \alpha_2)
\end{eqnarray*}

\item Given the generated model parameters and latent variables the pilot data $\mathbf{x}$ are then simulated from the PPCA model:
\begin{eqnarray*}
p(\underline{x}_{i}| \underline{u}_{i}, \textbf{W}, \sigma^2) &=& \mbox{MVN}_{p}(\textbf{W}\underline{u}_i, \sigma^2\textbf{I}) \:\: \mbox{for } i = 1, \ldots, n.
\label{eqn:model}
\end{eqnarray*}
\end{enumerate}
Estimating sample size based on pilot data simulated in this way ensures the estimated sample size is firmly dependent on the type of model being used to analyse the real experimental metabolomic data. Hence, MetSizeR represents an analysis based approach to sample size estimation for metabolomic studies. The specific steps involved in the MetsizeR algorithm are detailed in the next section.

\subsubsection*{The MetSizeR Algorithm}

The MetSizeR procedure for sample size estimation starts with a number $ntry$ of different sample sizes and a user-specified FDR (denoted by $target.fdr$). It then searches for the optimal sample size $\hat{n}$ by estimating the FDR for each of the $ntry$ sample sizes. In order to estimate FDR for each sample size, the null distribution of the test statistics of all metabolites is estimated and then a shift constant is added to the test statistics of some $p_o$ metabolites to allow them to be truly significant. The null distribution is estimated by calculating the test statistics of the permuted pilot data. After obtaining the critical values of the null distribution, the FDR is estimated. The optimal sample size $\hat{n}$ is then set to be the sample size with FDR equal to $target.fdr$.\\

In summary, the MetSizeR sample size estimation method proceeds as follows:

\begin{enumerate}
\item Specify the input parameters which include the desired level of FDR ($target.fdr$), the expected proportion $m$ of significant metabolites and the model to be used when analyzing the observed metabolomic data.
 \item Simulate pilot data of sample size $n_k$ from the assumed analysis model, where $k=1,\ldots,ntry$. Pilot data simulation from the PPCA model is detailed in the previous Section; the Additional File details pilot data simulation from the PPCCA and DPPCA models.
\label{item:data}
\item Estimate the null distribution for all metabolites by randomly permuting the group labels of the simulated pilot data and computing the test statistic $TS_{jt}$ for each metabolite $j$ and each permuted data set $t$ for $T$ permutations.
\item Estimate the FDR for each permuted data set $t=1,\ldots,T$:
\label{item:perm}
\begin{enumerate}
   \item Consider the corresponding $p$-vector of the test statistics $\underline{TS}_t=(TS_{1t},TS_{2t},\ldots, TS_{pt})$ for all metabolites on permutation $t$.
   \item Randomly sample $p_o = m \times  p$ of the test statistics $\underline{TS}_t$ and add $\frac{\delta}{\varrho_{jt}(\sqrt{\frac{1}{n_1}+\frac{1}{n_2}})}$ to their intensities. This allows $p_o$ metabolites to be truly significant. Here, $\delta$ is the effect size, and $\varrho_{jt}$ is the true within group standard deviation estimated by $\frac{S_{jt}}{\sqrt{\frac{1}{n_1}+\frac{1}{n_2}}}$.
   \item A cut off point $crit$ is set to be the $p_o^{th}$ largest absolute value of the test statistics $\underline{TS}_t$. All metabolites with $|TS_{jt}| > crit$ are declared as significant. The FDR for permutation $t$ can then be calculated.
\end{enumerate}
\item Estimate the FDR for data simulation $s$ by taking the $50^{th}$ percentile of the FDR values of $1,\ldots,T$ permutations.
\label{item:sim}
\item Repeat steps \ref{item:data} to \ref{item:sim} for $s=1,\ldots,SIM$ simulations and report the $10^{th}, 50^{th}$ and $90^{th}$ percentiles of the FDR values for sample size $n_k$.
\label{item:nk}
\item Repeat steps \ref{item:data} to \ref{item:nk} for $k=1,\ldots,ntry$ different sample sizes and select the optimal sample size $\hat{n}$ as the $n_k$ with FDR $=$ $target.fdr$.
\end{enumerate}

The total number of permutations $T$ used to estimate the sampling distribution of the test statistics $TS$ was chosen to be twenty. In the \texttt{samr} R package \cite{tibshirani06} 20 permutations were used to estimate the null distribution and they give accurate estimates of the FDR. Here, the value of the effect size $\delta$ is chosen based on the variance of the underlying model. The optimal sample size $\hat{n}$ is estimated by predicting the sample size at $target.fdr$ using a simple linear regression model on values of FDR above and below the $target.fdr$ with their corresponding sample sizes $n_k$. The estimated sample size by MetSizeR ensures that the power or the confidence level in statistical tests reaches (1-$target.fdr$).

\subsubsection*{Parameter Specification: details and guidelines.}

The MetSizeR algorithm requires the specification of several parameters; some are parameters relevant to the intended analysis model, and some are parameters relevant to the sample size estimation procedure itself.\\

In terms of the MetSizeR GUI which has been developed, the user is requested to specify parameters specific to the sample size estimation procedure i.e. the number of bins in the NMR or MS spectrum, the expected proportion of significant bins, the target FDR and the minimum sample size they wish to be considered. The default settings of these parameters are 200 spectral bins, 20\% significantly different bins, a target FDR of 5\% and a minimum sample size of 4. The choice of the number and proportion of significantly different spectral bins will naturally be informed by the metabolomic practitioner's knowledge, as will the minimum sample size choice. For the target FDR, again this depends on the conservatism of the metabolomic practitioner and/or the research question of interest, but a FDR of 5\% is indicative of typical statistical practice. The user can easily re-run the MetSizeR algorithm for different settings of these parameters to ascertain the influence of their particular specifications. However,
within the MetSizeR GUI the user has the option of requesting plots of the expected proportion of significant bins versus the FDR, over different sample sizes, giving insight to the influence of this particularly influential parameter on sample size estimation. Regarding the specification of parameters relevant to the intended analysis model, in the MetSizeR GUI, the user is only required to specify the intended analysis model (PPCA, PPCCA or DPPCA), and in the case of PPCCA, the number of covariates to be included. Both of these decisions are again practitioner informed, depending on the particular experiment under consideration. The MetSizeR manual, available through the developed MetSizeR GUI, guides the user through these parameter specification steps using a number of illustrative examples.\\

The remaining parameters in the MetSizeR algorithm have been fixed within the \textbf{R} code underlying the MetSizeR GUI, but given the open source nature of \textbf{R}, these can be changed by the user if desired. In the context of the PPCA model discussed above the hyperparameters of the prior distributions of the loadings matrix $\mathbf{W}$ and the variance $\sigma^{2}$ are based on previous estimates of $\mathbf{W}$ and $\sigma^2$ from applications of PPCA to metabolomic data (eg. \cite{nyamundanda10, nyamundanda12a}). Each row of the loadings matrix $\mathbf{W}$ is simulated from a standard multivariate Gaussian distribution MVN$_{q}(\underline{0}, \textbf{I})$ and the noise variance $\sigma^2$ is simulated from an inverse gamma distribution with shape parameter $\alpha_1=3$ and scale parameter $\alpha_2 = 4$. Hyperparameter settings for the PPCCA and DPPCA models are detailed in the Additional File. Within the MetSizeR algorithm four final parameters are specified: the effect size $\delta$ (fixed at 2.3, the 99th quantile of the
assumed prior distribution of the loadings), the correction factor $cf$ (fixed as the fifth percentile of the estimated standard errors of all metabolites), the number of permutations $T$ (set to 20) and the number of simulations $SIM$ (set at 20). These specifications are based on the choices in \cite{tibshirani06, lin10, Hwang02} in similar sample size estimation settings.\\

\section{Results }
This section illustrates the application of MetSizeR to different metabolomic experimental settings. In the first section, MetSizeR is employed to estimate sample size in the setting where experimental pilot data are not available; the second section considers the case where experimental pilot data are available.

\subsection{Sample size estimation using simulated pilot data}
\label{sec:pilot}

Here the MetSizeR approach to sample size estimation is illustrated in the setting where experimental pilot data are not available and it is assumed that the user has indicated that a PPCA model will be used to analyze the observed experimental data. Further, it is assumed that the user has specified that the spectra will consist of 300 spectral bins, the target FDR is 5\% and the expected proportion of significant spectral bins is 20\%. In this example, the user has also specified that they wish to consider a minimum sample size of ten, with five in each treatment group (i.e. $n_1=5$ and $n_2=5$). All other MetSizeR parameters are set at their default values, as detailed in the previous section. The MetSizeR method was then applied, and the $10^{th}, 50^{th}$ and $90^{th}$ percentiles of the FDR were calculated across a range of sample sizes and are shown in Figure 1. The sample size at which the target FDR of 5\% was achieved was estimated to be 30 with 15 in each treatment group as shown in Figure 1(A).\\

The expected proportion of significant spectral bins specified by the user impacts on the estimated number of samples required for the metabolomic experiment. Figures 1(B), 1(C) and 1(D) demonstrate the effect on FDR of varying the expected proportion of significant spectral bins for three different sample sizes. The figures show that, increasing the expected proportion of significant spectral bins reduces the FDR.\\

\begin{figure}[h!]
\begin{center}
\includegraphics[width=10cm, height=10cm]{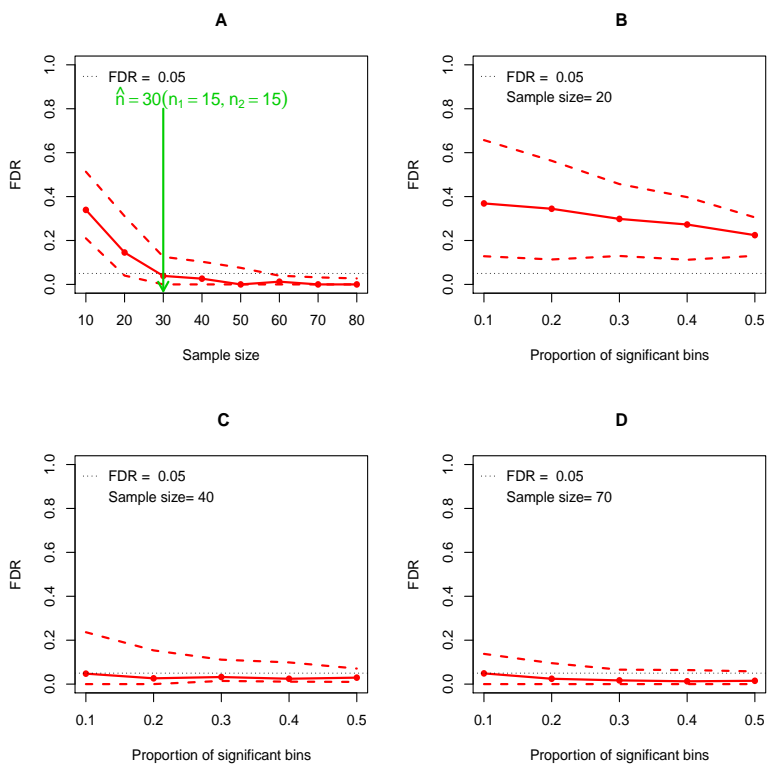}
\end{center}
\caption{In each panel is the estimated FDR (solid red lines) as well as the $10^{th}$ and $90^{th}$ percentiles (dashed red lines). A horizontal dashed black line is the target FDR at 5\%. \textbf{(A)} The sample size $\hat{n}$ is estimated to be 30 with 15 samples in each treatment group. \textbf{(B-D)} show the effect of varying the proportion of significant bins over a range of sample sizes.}
\end{figure}

A second example which demonstrates the applicability of MetSizeR is based on an experimental paradigm where additional information is available in the form of covariates. In this instance, the PPCCA model will be used to analyze the acquired data and thus was used to simulate pilot data with 300 spectral bins, five samples from each treatment group and two covariates. Fixing the target FDR at 5\% and the expected proportion at 20\%, Figure 2(A) demonstrates that when two covariates are included in the PPCCA model, the total number of samples required for such an experiment increases to 36 with 18 samples in each treatment group.\\

Figure 2(B) illustrates a third example of the setting where no experimental pilot data are available and the practitioner aims to conduct a longitudinal metabolomic experiment. The pilot data for this example are simulated from the DPPCA model; the data are simulated by only focusing on the first time point of the experiment as it is expected that the same number of subjects will be followed over time and that, while there may be dropouts, the largest number of subjects will be present at the first time point. Figure 2(B) shows that the expected number of samples required for a longitudinal study of 300 spectral bins with 20\% significant bins and a target FDR of 5\%, is 24 with 12 samples from each treatment group.

\begin{figure}[h!]
\begin{center}
\includegraphics[width=10cm, height=10cm]{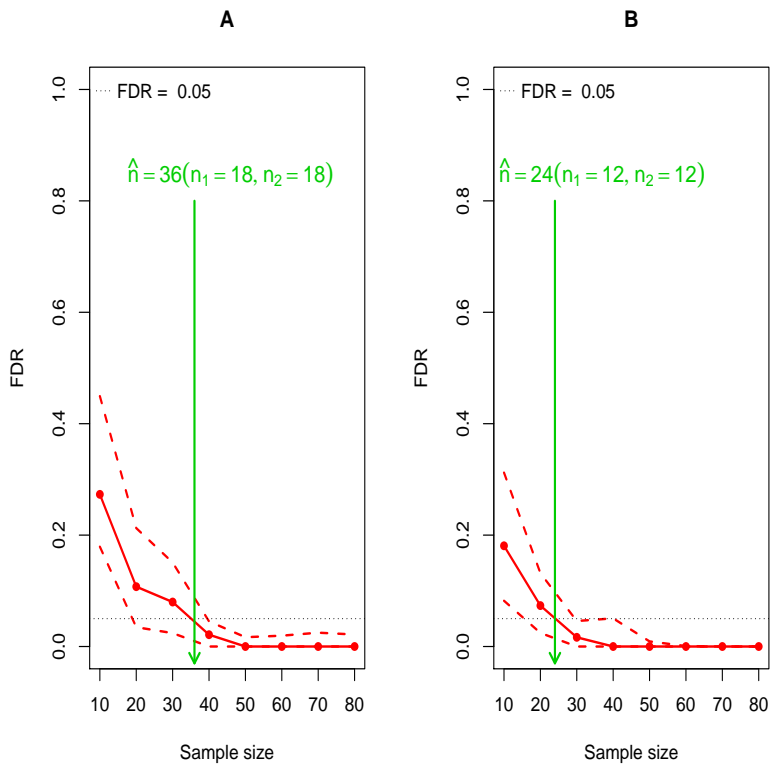}
\end{center}
\caption{\textbf{(A)} The estimated sample size using the PPCCA model with two covariates. \textbf{(B)} The estimated sample size for a longitudinal study using the DPPCA model.}
\end{figure}

\subsection{Sample size estimation with experimental pilot data}
\label{sec:pilot}

In a situation where experimental pilot data are available, parameter estimates used for simulations are based on fitting the underlying model to the experimental pilot data. Here, the application of MetSizeR is illustrated using real metabolomic data sets as experimental pilot data.\\

The first experimental pilot data set is from a longitudinal metabolomic animal study. Urine samples of 18 animals in two treatment groups were collected over a 15 day period and the animals' weights were measured. Details of this study have been previously detailed in \cite{Carmody10}. Data from day 10 of the study were used as experimental pilot data here; the NMR spectra consist of 189 spectral bins with nine samples in each treatment group. The PPCCA model was fitted to the experimental pilot data, with weight as a covariate and the maximum likelihood parameter estimates from fitting this model are used for data simulations in MetSizeR. Controlling the target FDR at 5\% and setting the expected proportion of significant bins at 20\%, the MetSizeR method was employed. Figure 3(A) depicts that the sample size estimate is 40, with 20 samples in each treatment group. It is interesting to note that, the 10\% and 90\% curves in Figure 3(A) are much narrower than in the previous examples in which MetSizeR was
used to estimate sample size with no experimental pilot data (Figures 1 and 2). This reduction in uncertainty is due to the fact that MetSizeR simulations are now based on fixed parameter values not on draws from prior distributions as used when experimental pilot data are not available.\\

The approach developed here for sample size estimation is not limited to NMR data. The method has been developed to accept data from targeted metabolomic analysis using MS, thus ensuring its applicability across the metabolomics community. Setting MetSizeR specifications as in the previous examples, the PPCA model was fitted to a targeted metabolomic MS pilot data set and under the MetSizR algorithm, the estimated sample size is shown in Figure 3(B).

\begin{figure}[h!]
\begin{center}
\includegraphics[width=10cm, height=10cm]{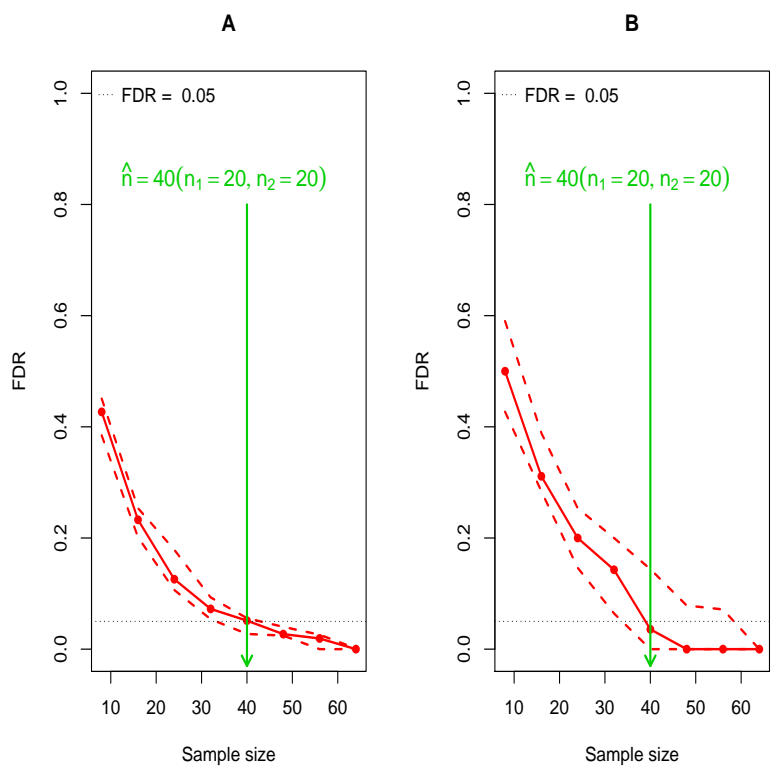}
\end{center}
\caption{\textbf{(A)} The estimated sample size using the PPCCA model on NMR pilot data with weights of subjects as a covariate. \textbf{(B)} The estimated sample size using the PPCA model with targeted MS metabolomic pilot data.}
\end{figure}

\section{Conclusions}
Determining sample sizes in metabolomics is important but due to the complexity of these experiments, there are currently no standard methods for sample size estimation in metabolomics. Moreover, since pilot studies are rarely done in metabolomics, sample size estimation approaches for high dimensional data studies requiring experimental pilot data, cannot be applied.\\

The method presented in this article is a straight forward approach for determining sample sizes for metabolomic experiments whilst controlling the FDR. The main advantage of the developed approach is its ability to determine sample size even when experimental pilot data are not available. Another key advantage is that it takes the type of analysis the researcher intends to use into consideration when estimating sample size and this can improve the power of the study. Also, since MetSizeR employs permutation techniques to estimate sample size, it accounts for correlation between metabolites and effect size variability. The method has been developed to accept both NMR and targeted MS data which will ensure wide applicability in the metabolomics community. Further, a software package facilitates easy implementation of the MetSizeR approach.\\

Areas of future work are multiple and varied. MetSizeR is currently designed to estimate the number of samples required for metabolomic experiments which involve two groups; modifications to the MetSizeR approach are possible to accommodate different metabolomic experimental designs. Alternatives to the permutation approach employed in MetSizeR could be examined -- bootstrap sampling would provide an interesting alternative. Proof of concept metabolomic experiments are currently underway to validate the MetSizeR approach.

 \bibliographystyle{Chicago}
  \bibliography{NyamundandaEtAl}

\end{document}